# Super-robust telecommunications enabled by topological half-supermodes

Rui Zhou[1,2†], Xintong Shi[1†], Hai Lin[1*], Yan Ren[1], Hang Liu[1], Zihao Yu[1], Jing Jin[1], Zhihao Lan[3] and Menglin L. N. Chen[2*†]

1. *College of Physical Science and Technology*, *Central China Normal University*, *Wuhan*, *Hubei 430079*, *China*.
2. *Department of Electrical and Electronic Engineering*, *The Hong Kong Polytechnic University*, *Hong Kong*, *China*.
3. *Department of Electronic and Electrical Engineering, University College London, London WC1E 6BT, U.K.*

[†] These authors contributed equally

[*]Correspondence to: Hai Lin and Menglin L. N. Chen

(e-mail: linhai@mail.ccnu.edu.cn, menglin.chen@polyu.edu.hk)

## Abstract

Topological photonics offers transformative potential for robust integrated waveguide devices due to their backscattering-immune properties. However, their integration faces two fundamental challenges: mode symmetry mismatch with conventional waveguides and prohibitive dimensions. We successfully overcome these two critical challenges by introducing a novel valley-ridge gap waveguide based on topological half-supermode engineering. By strategically hybridizing ridge waveguide modes and valley kink states, we create an exotic odd-symmetric supermode enabling robust propagation and ultra-compact operation. The further implementation of a perfect electric conductor boundary halves lateral dimensions while eliminating radiation loss. Crucially, our proposed valley-ridge interface achieves direct transverse electric mode matching with standard waveguides without transition structures, enabling seamless integration. Experimental results demonstrate reflection losses lower than -15 dB in realistic telecommunication scenarios with super-robust signal propagation through sharp bends. This work innovatively conceptualizes topological half-supermodes and pioneers their practical applications for integrated waveguide devices, establishing a completely new waveguide class that uniquely combines robust backscattering immunity with deep subwavelength compactness.

## Introduction

Waveguiding devices[1-2] form the foundation of modern high-frequency systems, supporting critical applications from 5G/6G networks to satellite communications and radar systems[3-5]. Among these, rectangular waveguides[6-7] are the most fundamental components that offer several key advantages over transmission lines, such as perfect shielding, low loss, and high-power capability. However, their fully enclosed metallic structure demands precise fabrication and makes them rigid, which limits their applications in compact and robust mmW systems.

The emergence of topological photonics[8-11], particularly valley Hall phases[12-17], has introduced new possibilities in backscattering-immune waveguiding against fabrication imperfections and sharp bends. Researchers have explored various integrated topological systems[18-29] by adopting the valley edge states in signal transportation. However, the valley edge states are inherently different from the fundamental modes in integrated systems, which brings difficulties in their interfacing with conventional devices. For example, the topological edge states are usually tightly confined at domain walls with non-adjustable mode width, making them not compatible with conventional waveguiding modes. Although the recently reported three-layer heterostructures[30-36] can introduce an additional width degree of freedom into topological waveguides, the topological modes being supported are local chiral states and require transit structures when matched to the conventional TEM or TE modes[36]. Moreover, the valley edge states with positive group velocity present asymmetric mode properties, which are incompatible with the symmetric fundamental modes in conventional waveguides[37-42]. Existing approaches to address this symmetry mismatch issue were based on complex transition structures, such as references [40-42]. However, it will inevitably introduce additional fabrication complexity and bandwidth restrictions. Meanwhile, valley photonic crystals (PCs) require periodic structures to create their bandgap properties, leading to increased lateral dimensions. Conventional miniaturization techniques, like in half-mode substrate-integrated waveguides (SIWs) [43], reduce the size in half but suffer from open-boundary radiation losses. These compounding limitations have prevented the realization of a unified platform that simultaneously achieves topological protection, compact dimensions, and a broadband interfacing strategy.

Compared with rectangular waveguides, gap waveguide (GW) technology[44-48] simplifies fabrication through non-contact EM confinement by using soft walls created by periodic structures. Inspired by the similarities between GWs and PCs, in this work, we leverage the valley Hall phases into GWs to provide robustness and highly compact routing. We propose a novel half-mode valley-ridge gap waveguide (VRGW) that overcomes the mode mismatch issues by leveraging two key innovations. Firstly, we generate an asymmetric topological hybrid supermode via controlled coupling between ridge modes and valley chiral states. By implementing a perfect electric conductor (PEC) boundary at the symmetry plane, a leakage-free 50% size reduction is achieved while the original mode profiles are preserved without any additional losses. Secondly, by tailoring the PC structures at the edge, the resultant half-topological hybrid supermodes enable seamless integration with conventional TE modes within a broad frequency band. Experimental validation across 24–27 GHz shows reflection losses below −15 dB, confirming efficient mode matching. Highly compact signal routing is also demonstrated by introducing sharp bends. By simultaneously achieving compact dimensions, leakage-free confinement, and broadband interfacing, our work establishes a new paradigm for high-performance mmW systems that bridges the gap between topological photonics and practical waveguide engineering.

## Results

**Design of the half-mode VRGW.** The proposed half-mode VRWG can be perfectly integrated into existing mmW systems with a compact size. As shown in Fig. 1a, the waveguide is connected with standard waveguide ports and the waveguiding path can be bent without reducing the transmission efficiency. This waveguide is designed and fabricated by adopting the GW technology at mmW frequencies, operating in the 24.5-27 GHz range, and its transverse dimensions are reduced by 50% compared to similar structures reported in references [36,40-42], as shown in Fig. 1b. Moreover, the waveguide manufacturing technology is fully compatible with computer numerical control (CNC) milling technology, possessing significant engineering application value, providing a reliable solution for the integrated design of the next generation of mmW systems.

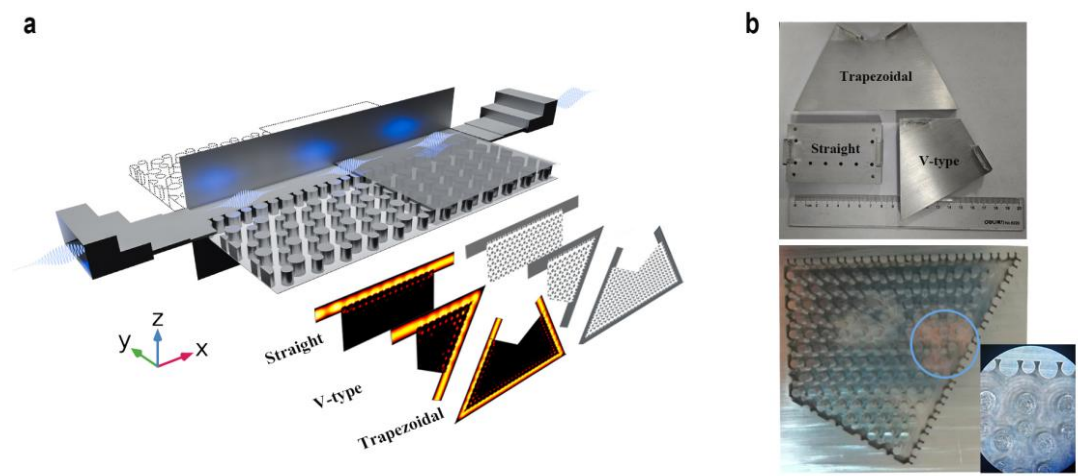

**Fig. 1:** Schematic of the half-mode VRGW. **a** Schematic of the proposed waveguide. It is connected with standard rectangular waveguide ports and three (straight, V-type, and trapezoidal) paths with unidirectional wave propagation are demonstrated. **b** Photos of the three fabricated samples and the fabrication details by CNC milling technology.

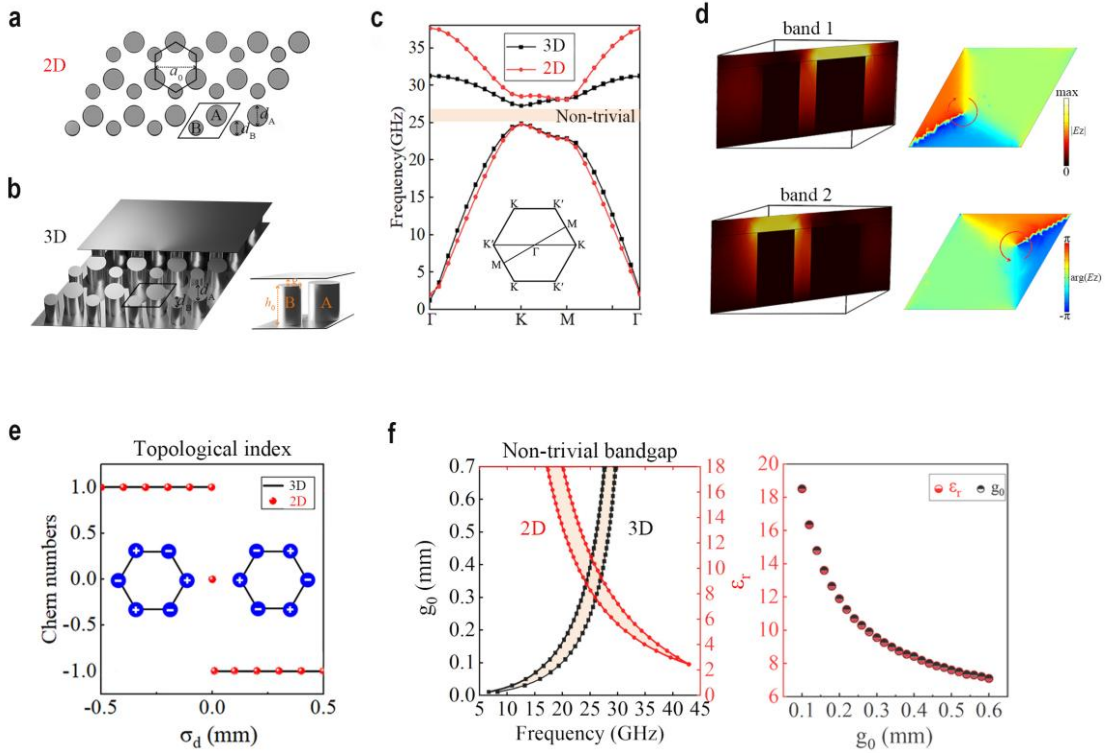

**Fig. 2:** Equivalence between the 2D dielectric PC and the 3D gapped metallic PC. **a** 2D dielectric PC. It is composed of two types of rods with diameters of $d_A$ and $d_B$. **b** 3D gapped metallic PC. It is composed of two types of finite-height pins with the diameters of $d_A$ and $d_B$. The pins are sandwiched between two metallic plates with the bottom one connected and the top one detached. The $h_0$ and $g_0$ are the heights of the pins and gaps, respectively. **c** Band structures of 2D dielectric PC (red line) and 3D gapped metallic PC (black line). **d** The amplitude and phase distributions of $E_z$ of the 3D gapped metallic PC at $K$-valley (bands 1 and 2). **e** The variations of the valley Chern number with respect to the difference ($\sigma_d = d_A - d_B$) between $d_A$ and $d_B$. The line and dots represent the results from the 3D gapped metallic PC and 2D dielectric PC, respectively. **f** The variations of the bandgap with respect to key parameters of the 2D dielectric PC ($\varepsilon_r$) and 3D gapped metallic PC ($g_0$) and their mappings.

**Valley topological phase transitions in metals and dielectrics.** To start with, the 2D valley PC is mimicked by using the GW technology. As shown in Figs.2a and b, a 2D dielectric PC and a 3D gapped metallic PC arranged in triangular lattices ($a_0$ = 3.5 mm) are simulated and compared. The dielectric rods have relative permittivities of 8.5. For the 3D gapped metallic PCs, the metallic pins have a limited height ($h_0$ = 1.85 mm). Note that the bottom of the pins are connected with the metallic plate, while there is a gap ($g_0$ = 0.38 mm) between the top of the pins and the metallic plate. To modulate the valley degree of freedom, the inversion symmetry of both the PCs are broken by varying the diameters of the rods and pins within the unit cells. Here, the size of the rods and pins are set to be the same and they are denoted by $d_A$ and $d_B$, respectively. As shown in Fig. 2c, the two valley PCs present a common bandgap between 25 and 27 GHz when $d_A$ = 1.7 mm, $d_B$ =1.2 mm. The eigenmodes of band 1 and band

2 at *K*-valley for the 3D gapped metallic PC are shown in Fig. 2d. Within the gap region of the 3D gapped metallic PC, the *K*-valley states of bands 1 and 2 demonstrate the same nature of mode profiles as in the 2D dielectric PC. The phase profiles of electric fields ($E$z) display distinct vortex characteristics, clearly revealing the presence of the non-zero valley Chern number. Hence, for the 2D dielectric PC and 3D metallic gapped PC, their natures of bulk properties present high consistency. The calculated valley Chern number for the difference ($\sigma_d = d_A - d_B$) between the sizes of the rods and pins are shown in Fig. 2e. The inset reveals topological indices $C_{K/K'} = \pm 1/2$ (**see Supplementary Note 1**). The variation in the valley Chern numbers $C_V = (C_K - C_{K'}) = \pm 1$ confirms the same topological phase transition of the two valley PC structures.

By changing the gap size $g_0$, the topological band boundaries of the 3D gapped metallic PC (black squares) can be regulated, which is similar to the trend of 2D dielectric PC (red points) by changing $\varepsilon_r$, as shown in Fig.2f. In the two valley PC structures, when the synchronous tuning condition ($f_{2D} = f_{3D}$) is satisfied, a significant spatial mapping characteristic emerges between the $g_0$ and $\varepsilon_r$. This mapping relationship enables the system to achieve controllable transformation of structural material properties while maintaining the integrity of the natural topological phase distribution. These results establish a theoretical bridge for material conversion in topological phase engineering, providing a technical platform for the design of mmW topological transmission devices.

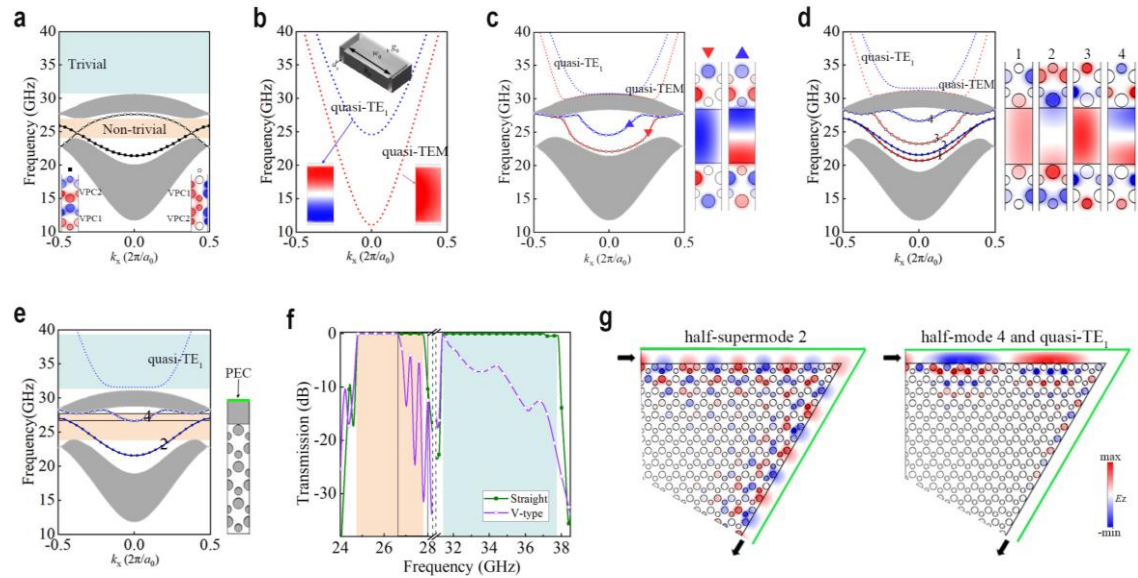

**Fig.3** Coupling principle of the topological hybrid half-supermodes. **a** Band structures and field distributions of the domain walls formed by VPC1 and VPC2. **b** Ridge waveguide and its two lowest eigenmodes, namely the quasi-TEM and quasi-TE$_1$ modes. By inserting a ridge at the domain walls of VPC1|VPC2 and VPC2|VPC1, the resulting band structures and field distributions are shown in **c** and **d,** respectively. **e** The band structure of the half-mode VRGW constructed by the PEC interface. **f** The transmission efficiency of the three modes in **e** with the straight and trapezoidal waveguide, and the field distribution in the V-type waveguide is shown in **g**.

**Topological hybrid half-supermodes in VRGW.** Robust transport is determined by the topological properties of the internal structure. We have demonstrated [16] that decorating the waveguide boundaries with 2D dielectric PC can protect the robust transmission of guide wave modes. Here, we utilize valley 3D gapped metallic PCs (VPCs) to decorate the boundaries of the ridge channel ($w_0 = 3.5l$), constructing a VRGW as shown in Fig. 3c. To investigate the EM behavior at the decorated boundary systematically, we split the VRGW into two parts: the valley gap PC waveguide and the ridge waveguide.

The valley gap PC waveguide is constructed with the VPC1 ($\sigma_d = 0.5$ mm) and VPC2 ($\sigma_d = -0.5$ mm) and its mirror-symmetry counterpart and presents the valley edge states within the non-trivial bandgap in Fig. 3a. The $E_z$ is localized at the VPC1|VPC2 and VPC2|VPC1 domain walls, with the positive and reverse group velocity, respectively. The ridge waveguide is designed to have two modes (quasi-TEM and quasi-TE$_1$ mode) within the range of the VPCs bandgap in Fig. 3b. The detailed parameters are: $h_0 = 1.85$ mm, $w_0 = 3.5l$, $g_0 = 0.38$ mm, $d_r = 0.625$ mm or 0.9 mm (**See Supplementary Note 3**). The quasi-TEM mode (red curve) field shows a symmetric distribution, while the quasi-TE$_1$ mode (blue curve) field shows an antisymmetric distribution along the y-axis. Since the edge state of VPC2|VPC1 is a backward transmission, it cannot effectively couple with the forward transmission mode of the ridge. Therefore, there are only two ridge modes that exist in the VRGW (within VPC2|VPC1), as shown in Fig. 3c. In the trivial bandgap, the dispersion of quasi-TEM and quasi-TE$_1$ modes in the ridge is still preserved. Due to the influence of the bulk state of the VPCs, these two modes have been relocated to the non-trivial bandgap, but have not had an effective coupling effect with the edge state (**See Supplementary Note 4**). The edge state of VPC1|VPC2 is a forward transmission mode. It couples with the forward transmission mode of the ridge, so that there are four modes in the non-trivial bandgap, as shown in Fig. 3d. From the lowest to highest eigenfrequencies at $k_x=0$, the four edge modes are denoted by modes 1 to 4. Modes 3 and 4 are the relocated ridge modes inside the non-trivial bandgap, which show similar mode fields as in Fig. 3c. Here, the edge modes 1 and 2 are topological hybrid supermodes formed by the coupling between the forward valley edge state and two ridge modes. The $E_z$ field distributions of supermodes reveal coupling characteristics, with the edge states and ridge modes exhibiting synchronous field information (more details **see Supplementary Note 5**).

Hybrid supermode 1 and mode 3 show the same symmetry as the quasi-TEM mode. Supermode 2 and mode 4 show the same symmetry as the quasi-TE$_1$ mode. They are divided into symmetric and asymmetric modes, which are well separated by building the boundaries between the VPC1 and a PMC/PEC. For supermode 1 and mode 3, the device size can be halved by using PMC simplification (**see Supplementary Note 6**). By inserting a PEC at the center of the ridge, only the asymmetric modes are preserved, as shown in Fig. 3e. By observing the transmission characteristics of the three half modes shown in e along the straight and trapezoidal (with two bends) paths, it is clear that the half-supermode 2 exhibits excellent stability during bending transmission due to its topological protection. In contrast, the half-modes 4 and quasi-TE$_1$ suffer significant energy loss in the bending due to the lack of topological protection, as shown in Fig. 3f-g.

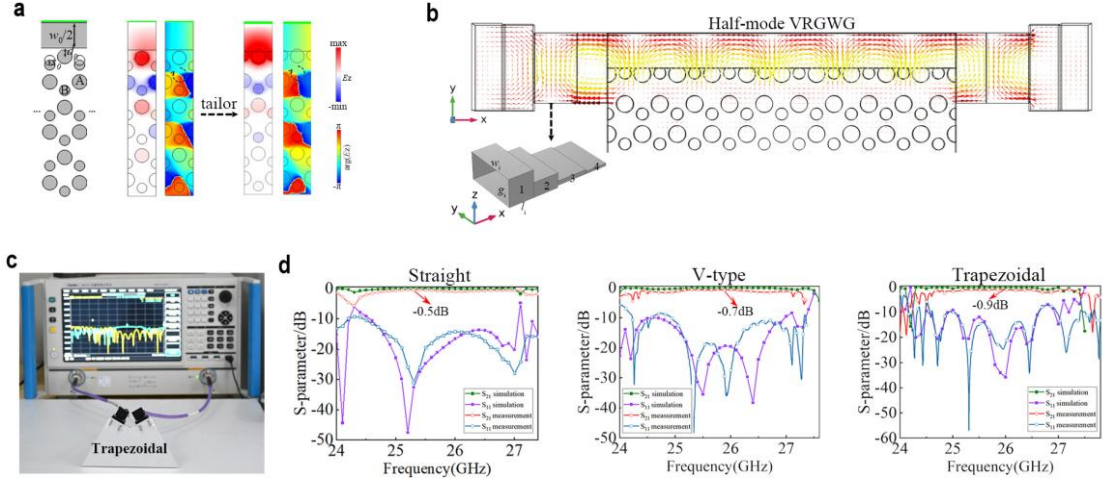

**Fig. 4** Full-wave simulation and experimental measurements on half-mode VRGW. **a** Schematic, field distributions, and band structures of half-mode VRGW, tailoring parameters ($w_0$= 3.5$l$, $d_0$ = ($l$-$d_A$)/2, $s_0$ = 1.1 mm). Where the $w_0$/2 signifies the width of the ridge, $d_0$ represents the distance from the ridge boundary to the A pins center, and $s_0$ indicates the displacement of the B pins. **b** Full-wave simulation of the straight half-mode VRGW with magnetic field line distribution. The subgraph is a three-view of the step-type rectangular waveguide, one end is connected to the standard waveguide WR34 (BJ260), and the other end is connected to the half-mode VRGW. **c** The Ceyear-3671G VNA is used to measure S-parameters of three waveguides and uses a WR34 (BJ260) to 2.4 mm female connector converter. **d** The simulated and measured S-parameters of the three half-mode VRGWs.

**Matching half-mode VRGW with standard components.** We construct a measurement platform compatible with an mmW integrated system. To suppress other interference modes, we use an tailoring half-mode VRGW ($w_0$= 3.5$l$, $d_0$ = ($l$-$d_A$)/2, $s_0$ = 1.1 mm) to keep only the topological hybrid half-supermode 2 in the non-trivial bandgap, as shown in Fig. 4a. Although the tailoring VPCs edge introduces system perturbations, its topological properties remain stable, which can be verified by the intact valley vortex phase. This stability is attributed to the inherent immunity of the valley topological protection mechanism to defects, indicating that the system has excellent fault tolerance to machining errors. In addition, the tailoring topological hybrid half-supermode 2 balances the coupling field between the quasi-TE$_1$ ridge mode and the edge state, thereby enhancing compatibility with standard components. According to the working frequency band, we select standard component WR34 (BJ260) for signal excitation and design a step-type rectangular waveguide for transition (**see**

**Supplementary Note 7**). The simulation shows that due to the mode field matching and boundary compatibility between the tailoring topological hybrid half-supermode and rectangular waveguide, the EM signal can be smoothly transmitted, which is clearly demonstrated by the magnetic field lines in Fig. 4b. The robust transmission of signals in diverse interference and bend paths remains a fundamental challenge in mmW integrated circuit design. Here, we have fabricated three distinct tailoring half-mode VRGWs for comprehensive empirical analysis: straight waveguide, V-type waveguide (with a 120-degree bend), and trapezoidal waveguide (with two 120-degree bends). The transmission efficiency, represented by the S-parameter, of three distinct half-mode VRGW types was evaluated using the vector network analyzer depicted in Fig. 4c. Experimental measurement data (Fig. 4d) indicate the insertion loss at the bend remains steadily at 0.2 dB. This phenomenon originates from the protection of the valley vortex phase, which effectively suppresses backscattering caused by waveguide bending, showing a significant difference from conventional waveguides (**see Supplementary Note 8**). The design provides a solution for the complex requirements of mmW systems, such as 5G base stations and radar.

**Verifications on telecommunication system.** The communication performance of the tailoring half-mode VRGW was experimentally validated by employing a universal software radio peripheral (USRP) platform. As illustrated in Fig. 5a, the experimental setup consists of a general-purpose computer serving as the master controller for both the USRP and UD Box frequency multiplier. The master computer transmitted data to the USRP via a PCI Express interface. The baseband signal was encoded using 16-QAM modulation in the USRP, followed by carrier up-conversion to 26 GHz through the UD Box frequency multiplier. The signals were then propagated in sequentially through three VRGWs for transmission. During the reception, the transmitted signal underwent down-conversion via the UD Box frequency multiplier and was sent back to the USRP. The data transmitted was obtained after demodulation.

Figure 5b illustrates the eye diagrams of three half-mode VRGWs linked into the telecommunication system: straight, V-type, and trapezoidal, respectively. All three samples demonstrated excellent signal integrity, showing consistent eye heights (1.4 V) and eye widths (6 μs), albeit with slight variations in noise levels. Meanwhile, the half-mode VRGWs are further validated with image transmission for a more intuitive demonstration. Fig. 5c depicts the received photographs through the three half-mode VRGWs. Additionally, real-time video transmission capability was experimentally validated, with demonstration provided in Supplementary Video 1 (**Telecommunication_experiment.mp4**). The topological phase protection mechanism endows the guided wave with robust transmission characteristics, enabling it to effectively traverse complex paths and imperfect structures while maintaining signal integrity and low loss. This robustness significantly enhances the system's reliability and lays a solid foundation for high-performance communication and signal processing.

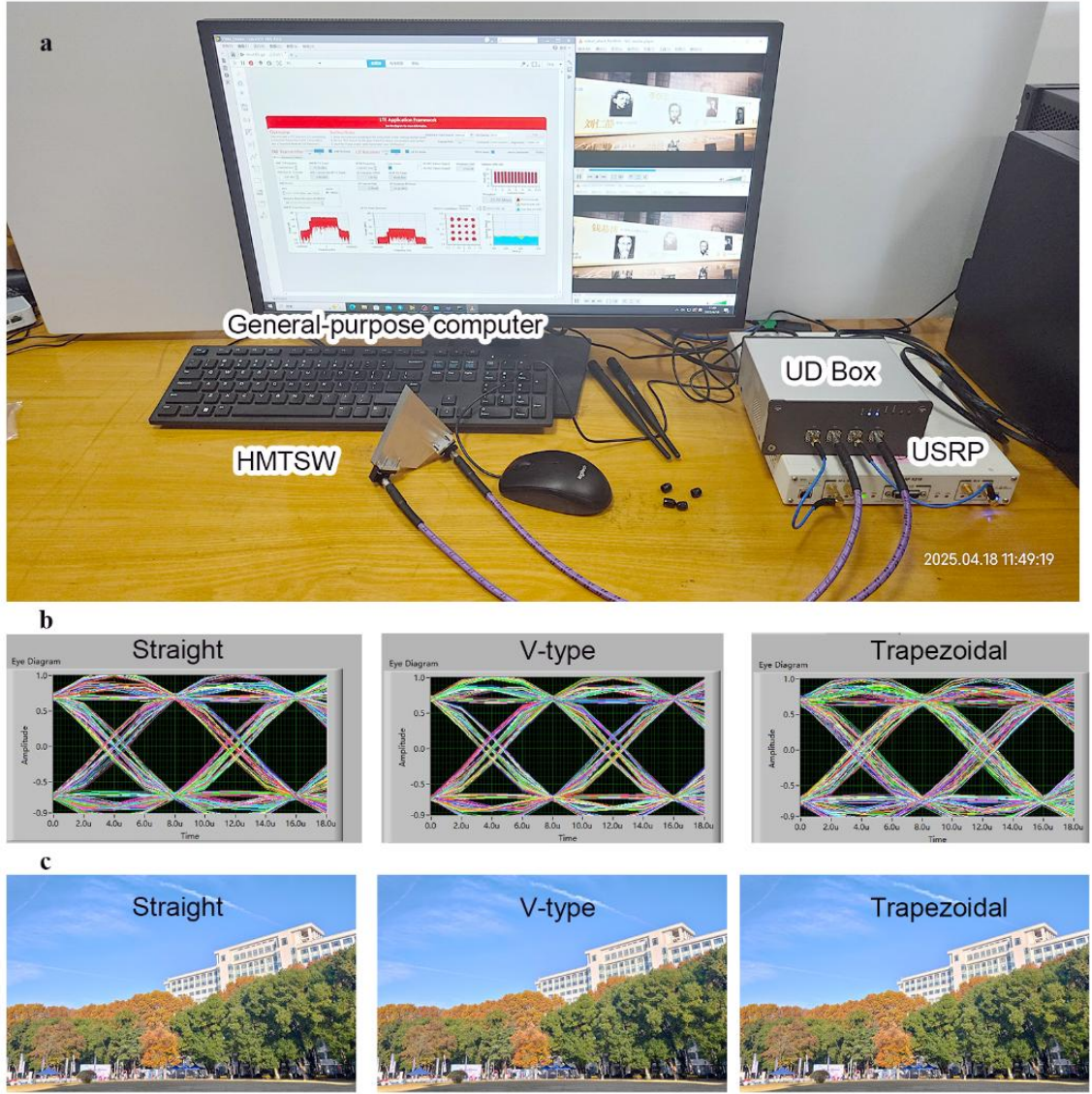

**Fig.5 Telecommunication experiments. a** The experimental scenario consists of a general-purpose computer serving as the master controller for both the USRP and UD Box frequency multiplier. **b** The eye diagrams of straight, V-type and trapezoidal tailoring half-mode VRGWs. **c** The original photograph and the received photographs through three half-mode VRGWs.

## Discussion

In conclusion, we have theoretically proposed and experimentally demonstrated a VRGW that fundamentally addresses two critical challenges in merging topological photonics with conventional waveguide systems. First, we resolve the inherent symmetry mismatch between topological edge states and conventional waveguide modes by engineering topological half-supermodes through controlled coupling of ridge-waveguide modes with valley kink states. Second, we eliminate excessive device dimensions via a PEC-truncated design, achieving 50% size reduction while preserving full mode confinement. All these features are essential for integrated waveguide systems. Measurements demonstrate direct TE$_{10}$-mode matching and robust backscattering suppression in three typical VRGW samples. This work successfully bridges the gap between topological photonics and waveguide engineering, enabling compact, high-performance telecommunication systems with inherent immunity to disorder and bends.

## Methods

**Sample fabrication.** The samples are composed of two polished aluminum plates. The bottom plate is precision-milled using a CNC machine to create a 3D VPC with specific lattice parameters, featuring dimensions of $d_A = 1.7$ mm, $d_B = 1.2$ mm, $a_0 = 3.5$ mm, and $h_0 = 1.85$ mm. The trapezoidal transition sections at both ends are also crafted on the bottom plate. The top plate is milled to have a groove structure with a height of $g_0 = 0.38$ mm.

**Numerical calculation.** The dispersion band calculation, eigenmode distribution, and EM field distribution shown in this work are calculated by a three-dimensional finite element method using the optical module of commercial software COMSOL MULTIPHYSICS.

**Experimental apparatus.** We employed a Vector Network Analyzer (VNA) [Ceyear-3671G] to measure the S-parameters of the waveguide. During the experiment, we initially calibrated the VNA using two 2.4 mm connecting cables to ensure the accuracy of the measurement results. Subsequently, two standard coaxial-to-waveguide adapters WR34 (BJ260) were mated and calibrated again to eliminate any transmission losses introduced by the WR34. After this step, the test waveguide was connected to the ports of the WR34, and S-parameter measurements were conducted over the specified frequency range. Throughout the measurement process, the VNA provided detailed information regarding the waveguide's reflection coefficient, transmission coefficient, and impedance characteristics. The software and hardware in telecommunication experiments are the National Instruments (NI) LTE Application framework, NI x310 USRP, and TMYTEK UD Box frequency multiplier.

## Contributions

All authors contributed extensively to the work presented in this paper. L. H*. and M.-L. C*† supervised all aspects of this work. R. Z† . and M.-L. C*† managed this project. R. Z† . carried out the theoretical analysis and experiments with the assistance of and X.-T. S†, M.-L. C*†and Z.-H. L. And H. L., Y. R., Z.-H. Y and J. J. put inputs together from all other coauthors in the paper revision.

## Data Availability

The authors declare no conflict of interest. The data that support the findings of this study are available from the corresponding author upon reasonable request.

# **Refercens**

## Acknowledgments

R. Z. and M. -L. C*. are supported by the National Natural Science Foundation of China Grant (62301470); Guangdong Basic and Applied Basic Research Foundation (2025A1515011622). R. Z., X. -T. S., H. L*., Y. R., Z.-H. Y., and J. J. are supported by National Natural Science Foundation of China (62301229); Ministry of Education Equipment Pre-Research Joint Fund under Grant (8091B032227); Self-Determined Research Funds of CCNU from the Colleges' Basic Research and Operation of MOE (CCNU23CG016); Nature Science Foundation of Hubei Province (2023AFB320) and Key Research and Development Project of Hubei Province (2023BAB061). We also thank Professor Song Han from Zhejiang University and Professor Yangjie Liu from Hubei University for their valuable insights that significantly enhanced this manuscript.